\documentstyle[seceq,preprint,epsbox]{jpsj}

\def\ve#1{\mbox{\boldmath $#1$}}
\def\runtitle{Stiffness of the Heisenberg Spin-Glass Model at Zero- and 
Finite-Temperatures in Three Dimensions}
\def\runauthor{Shin-ichi {\sc Endoh}, Fumitaka {\sc Matsubara} and 
Takayuki {\sc Shirakura}}

\title
{
Stiffness of the Heisenberg Spin-Glass Model at Zero- and 
Finite-Temperatures in Three Dimensions
}

\author
{ 
Shin-ichi {\sc Endoh}, Fumitaka {\sc Matsubara}
and Takayuki {\sc Shirakura}$^{1}$
}

\inst
{
Department of Applied Physics, Tohoku University, Sendai 980-8579\\
$^1$Faculty of Humanities and Social Science, Iwate University, 
Morioka 020-8550
}

\recdate
{
\today
}

\abst
{
 We examine the stiffness of the Heisenberg spin-glass (SG) model 
at both zero temperature ($T=0$) and finite temperatures ($T \ne 0$) 
in three dimensions.
 We calculate the excess energies at $T=0$ which are gained by rotating and 
reversing all the spins on one surface of the lattice, and find that they 
increase with the lattice size $L$. 
 We also calculate the excess free-energies at $T \ne 0$ which correspond 
to these excess energies, and find that they increase with $L$ at low 
temperatures, while they decrease with increasing $L$ at high temperatures.
 These results strongly suggest the occurrence of the SG phase at low 
temperatures.
 The SG phase transition temperature is estimated to be 
$T_{\rm SG} \sim 0.19J$ from the lattice size dependences of 
these excess free-energies.
}

\kword
{
Heisenberg spin-glass, stiffness, excess energy, excess free-energy, 
rotation, reversal
}

\begin{document}
\sloppy
\maketitle

\section{Introduction}
 Spin-glasses have attracted great challenge for computational physics.
 It has been believed that the spin-glass (SG) phase transition occurs
in three dimensions (3d) for the Ising model ~\cite{rf:1,rf:2,rf:3} but not 
for the XY and Heisenberg models.~\cite{rf:4,rf:5,rf:6,rf:7}
 However, numerical studies during the last decade have revealed that 
the SG phase might be more stable than what were previously believed.
 For example, in the isotropic Ruderman-Kittel-Kasuya-Yoshida (RKKY) model, 
it was suggested that the SG phase transition takes place 
at a finite temperature.~\cite{rf:8,rf:9}
 It was also suggested that, for the Ising model, 
the SG phase might be realized at low temperatures even 
in two dimensions (2d).~\cite{rf:10,rf:11,rf:12,rf:13}
 For the XY and Heisenberg SG models, Kawamura and coworkers took 
notice of the chirality described by three neighboring spins, and suggested 
that, in 3d, a chiral glass phase transition occurs at a finite temperature 
without the conventional SG order.~\cite{rf:14,rf:15,rf:16,rf:17,rf:18} 
 The basis of their suggestion is the stiffness of the system 
at zero temperature.~\cite{rf:14,rf:15}
 They estimated ${\theta}_c > 0$ and ${\theta}_s < 0$ from the lattice 
size dependences of the chiral and the spin domain-wall energies, where 
${\theta}_c$ and ${\theta}_s$ are the stiffness exponent of the chiralities 
and that of the spins, respectively.  
 Recently, however, the ambiguity was pointed out for calculating 
the domain-wall energy.~\cite{rf:12,rf:13,rf:19}
 In particular, Kosterlitz and Akino (KA) ~\cite{rf:19} calculated the spin 
domain-wall energy of the XY SG model using a different method and 
found that ${\theta}_s > 0$, contrary to Kawamura and coworker's 
suggestion.~\cite{rf:14}
 If KA's result is true, the question arises for the Heisenberg model 
whether the SG phase transition occurs at a finite temperature.

 The purpose of the present paper is to examine the stability of the SG 
phase in the Heisenberg model. 
 We consider the stiffness of the model at zero temperature ($T=0$) 
and finite temperatures ($T \ne 0$) calculating the excess energy and the 
excess free-energy, respectively, which are gained by rotating or reversing 
all the spins on one surface of the lattice.
 We find that the excess energy and the excess free-energy at low temperatures 
increase with the lattice size $L$.
 These results strongly suggest the occurrence of the SG phase at low 
temperatures.
 We estimate the transition temperature of $T_{\rm SG} \sim 0.19J$ from the 
lattice size dependence of the excess free-energy for various temperatures. 
 In section 2, after pointing out the ambiguity of the defect energy method 
conventionally used to estimate the domain-wall energy, we propose a method 
for considering the stiffness of the system.
 In section 3, we calculate the excess energy and discuss the stability 
of the SG phase at $T=0$.
 In section 4, we calculate the excess free-energy at $T \ne 0$ using the 
Monte Carlo twist method.~\cite{rf:20}
 Section 5 is devoted to the conclusions.
 Preliminary results of the present work were reported in Ref.~21.

\section{Model and Method}
 We start with the model described by the Hamiltonian
\begin{equation}
  {\cal H}=-\sum_{<ij>} J_{ij}{\ve{S}}_i \cdot {\ve{S}}_j
\end{equation}
where ${\ve{S}}_i$ is a Heisenberg spin with $|{\ve{S}}_i|=1$ and the sum
$<ij>$ runs over all nearest neighbor pairs.
 The coupling constant $J_{ij}$ takes $+J$ or $-J$ 
with the same probability $1/2$.

 Let us briefly describe the defect energy method conventionally used so far.
 The method originates from an application of the renormalization-group
concept.~\cite{rf:1,rf:4,rf:5}
 That is, one evaluates an effective coupling $\tilde{J}_{\rm eff}$ between 
block spins generated by renormalization.
 To estimate $\tilde{J}_{\rm eff}$, one considers the domain-wall energy 
$\Delta{E}$, which is usually defined as the difference in the ground state 
energy between two systems, A and B, with the same bond distribution but 
with different boundary conditions.
 For the system A, a periodic boundary condition is applied in every 
direction, and for the system B, an antiperiodic boundary condition is 
applied in one direction and the periodic boundary conditions in the 
other directions.
 However, the meaning of the domain-wall energy is not clear, because some 
domain-wall will arise in the system A and some different domain-wall will 
arise in the system B.
 Then, one might merely examine the difference in the energy between 
those domain-walls.
 It is also not clear whether the domain-wall energy $\Delta{E}$ really 
provides the effective coupling $\tilde{J}_{\rm eff}$ between the block spins. 

 Then, apart from the renormalization-group concept, we consider the 
stability of the spin configuration of the system itself.
 The strategy of our study is as follows.
 We prepare a cubic lattice of $L \times L \times (L+1)$ with an open 
boundary condition in one direction (z-direction) and the periodic 
boundary conditions in the other directions.
 That is, the lattice has two surfaces ${\Omega}_1$ and ${\Omega}_{L+1}$.
 First we determine the ground state spin configuration of the lattice.
 Hereafter, the ground state spin configurations on ${\Omega}_1$ and 
${\Omega}_{L+1}$ are denoted as $\{{{\ve{S}}_i}^{(1)}\}$ and 
$\{{{\ve{S}}_i}^{(L+1)}\}$, respectively.
 In this spin configuration, any distortion (domain-wall) will be 
removed, because the lattice has free surfaces ${\Omega}_1$ and 
${\Omega}_{L+1}$.
 Then we add a distortion inside the system in the manner that 
$\{{{\ve{S}}_i}^{(1)}\}$ are fixed and $\{{{\ve{S}}_i}^{(L+1)}\}$ are changed 
under the condition that the relative angles between the spins are fixed.
 The ground state energy of this system is always higher than 
that of the system before changing $\{{{\ve{S}}_i}^{(L+1)}\}$. 
 This excess energy is the net one added inside the system, 
because the surface energy of ${\Omega}_{L+1}$, which is given as the sum of 
the exchange energies between the spins on ${\Omega}_{L+1}$, is conserved.
 We consider the stability of the system on the basis of this excess energy. 
 One might think that the fixing of the relative spin directions on 
${\Omega}_1$ and ${\Omega}_{L+1}$ overestimates the stability of the spin
configuration.
 We think, however, that this restriction is not serious for discussing
the stability, because the increase of the excess energy to infinity for
$L \to \infty$ means nothing but the existence of a strong correlation 
between the spin configurations on ${\Omega}_1$ and ${\Omega}_{L+1}$.
 In fact, the same method was successfully applied to the Ising SG model
in 2d.~\cite{rf:12,rf:13}

\section{The Stiffness at ${\bf T=0}$}
 In this section, we consider the stiffness of the system at $T=0$ 
calculating two kinds of excess energies. 
 One is the excess energy which is gained by rotating 
$\{{{\ve{S}}_i}^{(L+1)}\}$ and the other is the excess energy which is 
gained by reversing $\{{{\ve{S}}_i}^{(L+1)}\}$.
 We think that it is sufficient to examine these two excess energies for 
considering the stiffness, because we can change $\{{{\ve{S}}_i}^{(L+1)}\}$ 
into any direction by combining the rotation and the reversal. 

 First we consider the stiffness of the system when we rotate 
$\{{{\ve{S}}_i}^{(L+1)}\}$.
 The spins $\{{{\ve{S}}_i}^{(L+1)}\}$ are rotated around a common axis 
by the same angle $\phi$.
 Here we note that effective rotating angles of the spins on ${\Omega}_{L+1}$ 
are not the same, because $\{{{\ve{S}}_i}^{(L+1)}\}$ have different 
directions even in the ground state.
 That is, the spin parallel to the axis does not rotate, whereas the spin 
perpendicular to the axis rotates by the angle $\phi$.
 Then, as a measure of the rotating angle, we introduce 
the averaged rotation angle $\omega$ defined by
\begin{equation}
  \cos\omega=\frac{1}{L^2}\sum_{i}{{\ve{S}}_i}^{(L+1)}
             \cdot{{\ve{S}}_i}^{(L+1)}(\phi),
\end{equation}
where $\{{{\ve{S}}_i}^{(L+1)}(\phi)\}$ is the set of the spins 
on ${\Omega}_{L+1}$ after the rotation.
 Hereafter, we call the system with the surface spin configurations 
$\{{{\ve{S}}_i}^{(1)}\}$ and $\{{{\ve{S}}_i}^{(L+1)}\}$ the reference system 
and that with the surface spin configurations $\{{{\ve{S}}_i}^{(1)}\}$ and 
$\{{{\ve{S}}_i}^{(L+1)}(\phi)\}$ the rotated system. 
 An important point is that there is an upper bound of $\omega$ which 
can be reached by this rotation.
 If $\{{{\ve{S}}_i}^{(L+1)}\}$ are randomly distributed, then the 
maximum value ${\omega}_{\rm max}$ for $L \to \infty$ is given as
\begin{eqnarray}
  \omega_{\rm max} &=& \lim_{L \to \infty} \cos^{-1} \Bigl(\sum_{i}
                       {{\ve{S}}_i}^{(L+1)} \cdot
                       {{\ve{S}}_i}^{(L+1)}(\pi) \Bigl) \nonumber \\
                   &=& \cos^{-1} \Bigl(\frac{1}{4\pi}\int_0^{2\pi}d\phi
                       \int_0^{\pi}(\cos^2\theta-\sin^2\theta)\sin\theta
                       d\theta \Bigl) \nonumber \\
                   &=& \cos^{-1}\Bigl(-\frac{1}{3}\Bigl)~\sim~0.6\pi.
\end{eqnarray}
 Therefore, it will be sufficient for us to consider the stiffness for 
$\omega \le {\omega}_{\rm max}$, because we consider the stiffness for 
$L \to \infty$.
 In fact, the maximum value of $\omega$ decreases as the number of the spins 
on ${\Omega}_{L+1}$ is increased (see Fig.~\ref{fig:1}).
 
 The excess energy $\Delta{E}_{\rm rot}(\omega)$ due to the rotation may 
be given as
\begin{equation}
  \Delta{E}_{\rm rot}(\omega)=E_{\rm rot}(\omega)-E_G,
\end{equation}
where $E_G$ and $E_{\rm rot}(\omega)$ are the ground state energy of the 
reference system and the lowest energy of the rotated system with $\omega$,
respectively.
 If $\Delta{E}_{\rm rot}(\omega)$ for $\omega \le {\omega}_{\rm max}$ 
increases with the lattice size $L$, the ground state spin configuration 
will be stable.
 Otherwise, the spin configuration is unstable.
 This concept is, of course, justified in the ferromagnetic Heisenberg model 
in $d$-dimensions, that is, 
$\Delta{E}_{\rm rot}(\omega)=\frac{1}{2}J{\omega}^2L^{\theta}$,
where $\theta~(=d-2)$ is the stiffness exponent. 

 In this calculation, it is not easy to obtain the correct value of 
$E_{\rm rot}(\omega)$, because there are infinite rotating axes which 
give the same angle $\omega$.
 Here we estimate $E_{\rm rot}(\omega)$ as follows.
 For each $\omega$, we prepare $N_r(=200)$ rotating axes and calculate 
the minimum energy for each of those axes.
 The energy $E_{\rm rot}(\omega)$ is approximated by the lowest value 
of those minimum energies.
 Of course, $E_{\rm rot}(\omega)$ is not the correct one but an upper bound 
of it, because $N_r$ is finite.
 In fact, $E_{\rm rot}(\omega)$ for $N_r/2$ is slightly larger than that 
presented here (see Fig.~\ref{fig:2}).
 However, we think $\Delta{E}_{\rm rot}(\omega)$ obtained here gives 
qualitatively correct $\omega$- and $L$- dependences, because the data 
for $N_r$ and those for $N_r/2$ yield almost the same values of exponents 
$\alpha$ and ${\theta}_{\rm rot}$ defined below.

 The lattice sizes studied here are $L=3 \sim 8$ and, for each $L$, the sample 
averages are taken over about 1000 independent bond realizations.
 In the calculation of $E_G$ and $E_{\rm rot}(\omega)$, we prepare about 
1000 initial spin configurations, which is obtained by annealing the system 
down to $T=0.1J$ from a higher temperature, and apply the spin-quench 
algorithm.~\cite{rf:4,rf:5}

 In Fig.~\ref{fig:1}, we present the results of the $\omega$-dependence of
$[\Delta{E}_{\rm rot}(\omega)]$ for various sizes of $L$, where 
$[\cdot\cdot\cdot]$ means the sample average.
 We find that for $\omega \le {\omega}_{\rm max}$ 
$[\Delta{E}_{\rm rot}(\omega)]$ increases with $L$, implying that 
the ground state is stable against the rotation.
 To confirm this view, in Fig.~\ref{fig:2}, we plot 
$[\Delta{E}_{\rm rot}(\omega)]$ as functions of $L$ for various $\omega$
in a log-log form.
 All the data for each $\omega$ lie on a straight line with almost the same 
slope 0.8, that is, 
$[\Delta{E}_{\rm rot}(\omega)] \propto L^{{\theta}_{\rm rot}}$ with 
${\theta}_{\rm rot}=0.8 \pm 0.1$ for $\omega \le {\omega}_{\rm max}$.
 We also examine the $\omega$-dependence of the coefficient and find that 
the data for $\omega \le {\omega}_{\rm max}$ fit the relation
\begin{equation}
  [\Delta{E}_{\rm rot}(\omega)] \sim A{\omega}^{\alpha}
                                     L^{{\theta}_{\rm rot}},
\end{equation}
with $\alpha \sim 1.9$, ${\theta}_{\rm rot} \sim 0.8$, and $A \sim 0.14J$.
 This finding is illustrated by the scaling plot in Fig.~\ref{fig:3}.
 
 This result is quite interesting, because it suggests that the SG phase is 
realized at low temperatures.
 A surprising point is that the values of $\alpha$ and ${\theta}_{\rm rot}$ 
are close to those of $\alpha = 2$ and $\theta = 1$ of the ferromagnetic 
Heisenberg model in 3d, which implies that the response of the spin 
configuration of the SG model against the rotation is analogous to that 
of the ferromagnetic model.
 In fact, we found that the rotated angles of the spins gradually 
increase from layer to layer, reminicent of the spin-wave 
excitation.~\cite{rf:21}
 We note, however, that the coefficient of $A \sim 0.14J$ is much smaller 
than that of $A=0.5J$ in the ferromagnetic model. 

 Next we consider the stiffness of the system when we reverse 
$\{{{\ve{S}}_i}^{(L+1)}\}$.
 Hereafter, we call the system with the surface spin configurations 
$\{{{\ve{S}}_i}^{(1)}\}$ and $\{-{{\ve{S}}_i}^{(L+1)}\}$ the reversed system.
 The excess energy $\Delta{E}_{\rm rev}$ due to the reversal 
may also be given as 
\begin{equation}
   \Delta{E}_{\rm rev}=E_{\rm rev}-E_G,
\end{equation}
where $E_{\rm rev}$ is the ground state energy of the reversed system.
 It should be noted here that $\{-{{\ve{S}}_i}^{(L+1)}\}$ is never realized 
by any rotation of $\{{{\ve{S}}_i}^{(L+1)}\}$.~\cite{rf:15}
 Thus $\Delta{E}_{\rm rev}$ will give a different view on the stability of 
the system.
 The sample averages are taken over about 4000 $(L=3 \sim 6)$ and 2000 
$(L=7,8)$ independent bond realizations.
 The result of the $L$-dependence of $[\Delta{E}_{\rm rev}]$ is presented 
in Fig.~\ref{fig:4}.
 We also find that $[\Delta{E}_{\rm rev}]$ increases with $L$, suggesting 
that the ground state is stable against the reversal.
 However, it is difficult to estimate the value of the stiffness exponent 
${\theta}_{\rm rev}$ for the reversal, because the slope of the curve 
gradually increases with $L$.
 Here, we tentatively estimate it as ${\theta}_{\rm rev} = 0.4 \pm 0.1$ 
from the data for $L=6,7,8$.
 It should be emphasized that this value might not be the one for 
$L \to \infty$ but a lower bound of it.

 These results of the stiffness for both the rotation and the reversal 
strongly suggest the occurrence of the SG phase transition 
at a finite temperature.
 It is noted, however, the estimated value of ${\theta}_{\rm rev} \sim 0.4$ 
is considerably smaller than that of ${\theta}_{\rm rot} \sim 0.8$.
 There is a speculation that ${\theta}_{\rm rot}$ could not be larger than 
${\theta}_{\rm rev}$.~\cite{rf:22}
 Could ${\theta}_{\rm rev}$ be equal to ${\theta}_{\rm rot}$ for 
$L \to \infty$ ?
 In fact, the result of $[\Delta{E}_{\rm rev}]$ shown in Fig.~\ref{fig:4} 
implies the increase of ${\theta}_{\rm rev}$ with $L$.
 To examine this possibility, we also calculate the defect energies 
$[\Delta{E}_{\rm def}]$ of the system using replica boundary 
conditions,~\cite{rf:23} where $\Delta{E}_{\rm def}=E_R-E_G$.
 That is, we prepare two replicas with the same bond distribution and two 
surfaces $S_1$ and $S_2$.
 The surfaces $S_1$ of the two replicas are connected with each other by the 
ferromagnetic bond of $(J,J,J)$, where $(J,J,J)$ is the set of the 
connected bonds in the three spin components.
 We calculate the ground state energy $E_G$ of the system under the condition 
that the other surfaces $S_2$ are also connected by $(J,J,J)$.
 Then we calculate $E_R$ under two different boundary conditions for the 
surfaces $S_2$.
 In the case A, they are connected by the antiferromagnetic 
bond of $(-J,-J,-J)$, which may correspond to the reversal of 
$\{{{\ve{S}}_i}^{(L+1)}\}$ in the present model.
 In the case B, they are connected by an asymmetric bond of 
$(-J,-J,J)$, which may correspond to the rotation around the z-axis by
$\phi = \pi$.
 These results are presented in Fig.~\ref{fig:5}.
 We find that, the defect energies in both cases increase with $L$ and their 
slopes seem to approach as $L$ is increased.
 However, further studies are necessary to examine the possibility of 
${\theta}_{\rm rev} = {\theta}_{\rm rot}$ for $L \to \infty$.
 Anyway, all the results of the stiffness examined here suggest the 
occurrence of the SG phase at low temperatures.

\section{The Stiffness at ${\bf T \ne 0}$}
 In this section, we consider the stiffness of the system at $T \ne 0$ to 
estimate the SG phase transition temperature $T_{\rm SG}$.
 We examine two kinds of excess free-energies which correspond to the 
excess energies $\Delta{E}_{\rm rot}(\omega)$ and $\Delta{E}_{\rm rev}$ 
examined in ${\S}$ 3.
 First we prepare the reference system in which the spin configurations 
$\{{{\ve{S}}_i}^{(1)}\}$ and $\{{{\ve{S}}_i}^{(L+1)}\}$ on ${\Omega}_1$ and 
${\Omega}_{L+1}$ are fixed to those of the ground state.
 Then we calculate the excess free-energies $\Delta{F}_{\rm rot}(T)$ and 
$\Delta{F}_{\rm rev}(T)$ which are gained by rotating and reversing 
$\{{{\ve{S}}_i}^{(L+1)}\}$, respectively.
 Here, we consider $\Delta{F}_{\rm rot}(T)$ when $\{{{\ve{S}}_i}^{(L+1)}\}$ 
is rotated around a fixed axis (the z-axis) by $\pi/2$, because it is not 
easy to find the rotating axis which gives the minimum free-energy of 
the rotated system.
 We calculate $\Delta{F}_{\rm rot}(T)$ and $\Delta{F}_{\rm rev}(T)$ using 
the Monte Carlo twist method proposed by Ueno.~\cite{rf:20}
 These excess free-energies are defined as
\begin{equation} 
   \Delta{F}_{\rm rot,rev}(T)=F_{\rm rot,rev}(T)-F_G(T),
\end{equation}
where $F_G(T)$ is the free-energy of the reference system, and 
$F_{\rm rot}(T)$ and $F_{\rm rev}(T)$ are the free-energies of the 
rotated and reversed systems, respectively.
 We may rewrite eq.(4.1) as
\begin{equation}
   \Delta{F}_{\rm rot,rev}(T)=\Delta{E}_{\rm rot,rev}(T)
                             -T\Delta{S}_{\rm rot,rev}(T),
\end{equation}
where $\Delta{E}_{\rm rot}(T)$ and $\Delta{S}_{\rm rot}(T)$ are the 
differences in the internal-energy and the entropy between the reference 
system and the rotated system, respectively, and $\Delta{E}_{\rm rev}(T)$ and 
$\Delta{S}_{\rm rev}(T)$ are those between the reference system and the 
reversed system.
 Here $\Delta{E}_{\rm rot}(T)$ and $\Delta{E}_{\rm rev}(T)$ are obtained 
by using the conventional Monte Carlo method.  
 On the other hand, $\Delta{S}_{\rm rot}(T)$ and $\Delta{S}_{\rm rev}(T)$ 
are obtained by means of the numerical integration of 
$\Delta{E}_{\rm rot}(T)$ and $\Delta{E}_{\rm rev}(T)$ from a high 
temperature, that is,
\begin{equation}
   \Delta{S}_{\rm rot,rev}(T)=\int_{T_0}^{T}
                              \frac{d\Delta{E}_{\rm rot,rev}(T^{'})}{T^{'}},
\end{equation}
where $T_0$ is a temperature high enough to satisfy 
$\Delta{E}_{\rm rot}(T_0)=0$ and $\Delta{E}_{\rm rev}(T_0)=0$.

 The lattice sizes studied here are $L=4 \sim 10$ and, for each $L$, the 
sample averages are taken over about 2000 independent bond realizations.
 The starting temperatures are $T_0/J = 3.0~(L=4)$, $T_0/J = 2.1~(L=5)$, 
$T_0/J = 1.7~(L=6)$, $T_0/J = 1.5~(L=7)$, $T_0/J = 1.3~(L=8)$, 
$T_0/J = 1.1~(L=9)$ and $T_0/J = 1.0~(L=10)$.
 We use a gradual-cooling method and take $4 \sim 5 \times 10^4$ MCS for the 
calculation of $\Delta{E}_{\rm rot}(T)$ and $\Delta{E}_{\rm rev}(T)$.  

 In Figs.~\ref{fig:6} and ~\ref{fig:7}, we show the temperature dependences 
of $[\Delta{F}_{\rm rot}(T)]$ and $[\Delta{F}_{\rm rev}(T)]$, respectively.
 The data for $T=0$ obtained by using the method in ${\S}$ 3 are also 
plotted in these figures.   
 We find that, at high temperatures $[\Delta{F}_{\rm rot}(T)]$ and 
$[\Delta{F}_{\rm rev}(T)]$ decrease with increasing $L$, whereas at low 
temperatures they increase with $L$.
 These results suggest that the SG phase will be realized at low temperatures.
 Usually, one estimates the phase transition temperature from the crossing 
temperature of the free-energies for various lattice sizes $L$.
 In the present model, however, we could not use this method directly, 
because the crossing temperature $T_L$ for the lattice sizes $L$ and $L+1$ 
shifts systematically to the low temperature side with increasing $L$.
 Then, we tentatively assume that $T_L$'s for $[\Delta{F}_{\rm rot}(T)]$ and 
$[\Delta{F}_{\rm rev}(T)]$ decrease linearly with $1/L$, and plot them as 
functions of $1/L$, which are shown in Fig.~\ref{fig:8}.
 In fact, $T_L$ for each of the cases seems to lie on a straight line.
 Using the least-squares method, we estimate $T_{\infty}/J = 0.192 \pm 0.015$ 
for the rotation and $T_{\infty}/J = 0.188 \pm 0.015$ for the reversal.
 A remarkable point is that these values of $T_L$ for $L \to \infty$ are 
almost the same.
 Therefore we may conclude that, if the SG phase transition really occurs, 
the transition temperature should be $T_{\rm SG} \sim 0.19J$.

\section{Conclusions}
 We have examined the stiffness of the Heisenberg SG model in 3d 
at both $T=0$ and $T \ne 0$, using a method which concerns with 
the stability of the spin configuration of the system itself.
 We calculated the excess energies $[\Delta{E}_{\rm rot}(\omega)]$ and 
$[\Delta{E}_{\rm rev}]$ at $T=0$ which are gained by rotating all the spins 
on one surface of the lattice and reversing them, respectively, and found that 
both $[\Delta{E}_{\rm rot}(\omega)]$ and $[\Delta{E}_{\rm rev}]$ increase 
with the lattice size $L$. 
 This result strongly suggests the occurrence of the SG phase transition 
at $T_{\rm SG} \ne 0$.
 Then, we calculated the excess free-energies $[\Delta{F}_{\rm rot}(T)]$ and 
$[\Delta{F}_{\rm rev}(T)]$ at $T \ne 0$ to estimate the transition 
temperature.
 We found that each of these free-energies increases with $L$ at low 
temperatures, whereas it decreases with increasing $L$ at high temperatures.
 From these results, we estimated the SG phase transition temperature 
as $T_{\rm SG} \sim 0.19J$.
 Recently we also performed the Monte Carlo simulation and found an evidence 
of the SG phase transition at almost the same temperature of 
$T_{\rm SG} \sim 0.18J$.~\cite{rf:24}
 Hence we believe that the SG phase is really realized in this model and the 
transition temperature to this phase is $T_{\rm SG}/J = 0.19 \pm 0.02$. 
 However, further studies are required to confirm our view.
 We hope that the present work stimulates other works on vector spin glasses.

\section*{Acknowledgements}
 The authors would like to thank Professor K. Sasaki, Dr. T. Nakamura, 
Professor H. Takayama, Professor Y. Ueno, and Dr. Y. Ozeki for their 
useful discussions.
 Numerical calculation was mainly performed on SX-4 at the supercomputer
center, Tohoku University.

\newpage

\begin{figure}
%  \psbox[scale=0.9]{fig1}
  \caption{The excess energies $[\Delta{E}_{\rm rot}(\omega)]$ of the 
           Heisenberg SG model in 3d for various lattice sizes of $L$ as 
           functions of the rotating angle $\omega$. The arrow indicates 
           the position of ${\omega}_{\rm max}$.}
  \label{fig:1}
%\end{figure}
%\newpage
%\begin{figure}
%  \psbox[scale=0.9]{fig2}
  \caption{The lattice size dependences of the excess energies 
           $[\Delta{E}_{\rm rot}(\omega)]$ of the Heisenberg SG model in 3d 
           for various rotating angles $\omega$. Cross and circle represent 
           the data for 100 and 200 rotating axes, respectively.}
  \label{fig:2}
%\end{figure}
%\newpage
%\begin{figure}
%  \psbox[scale=0.9]{fig3}
  \caption{Scaling plots of the excess energies 
           $[\Delta{E}_{\rm rot}(\omega)]$ of the Heisenberg SG model in 3d. 
           Here ${\theta}_{\rm rot} = 0.8$ and $\alpha = 1.9$. 
           We show the results only in the $\omega$- range which is relevant 
           in the model (the arrow indicates the position of 
           ${\omega}_{\rm max}$).}
  \label{fig:3}
%\end{figure}
%\newpage
%\begin{figure}
%  \psbox[scale=0.9]{fig4}
  \caption{The lattice size dependence of the excess energy 
           $[\Delta{E}_{\rm rev}]$ of the Heisenberg SG model in 3d.}
  \label{fig:4}
%\end{figure}
%\newpage
%\begin{figure}
%  \psbox[scale=0.9]{fig5}
  \caption{The lattice size dependences of the defect energies of the 
           Heisenberg SG model in 3d which are calculated by using two 
           replica boundary conditions explained in the text.}
  \label{fig:5}
%\end{figure}
%\newpage
%\begin{figure}
%  \psbox[scale=0.9]{fig6}
  \caption{Temperature dependences of the excess free-energies 
           $[\Delta{F}_{\rm rot}(T)]$ of the Heisenberg SG model in 3d.
           Note that $[\Delta{F}_{\rm rot}(T)]$ at $T=0$ is the excess 
           energy $[\Delta{E}_{\rm rot}]$ which is gained by rotating 
           around the z-axis by $\pi/2$, and it is calculated by the 
           different method in ${\S}$ 3.}
  \label{fig:6}
%\end{figure}
%\newpage
%\begin{figure}
%  \psbox[scale=0.9]{fig7}
  \caption{Temperature dependences of the excess free-energies 
           $[\Delta{F}_{\rm rev}(T)]$ of the Heisenberg SG model in 3d.
           Note that $[\Delta{F}_{\rm rev}(T)]$ at $T=0$ is the excess 
           energy $[\Delta{E}_{\rm rev}]$ calculated by the different 
           method in ${\S}$ 3.}
  \label{fig:7}
%\end{figure}
%\newpage
%\begin{figure}
%  \psbox[scale=0.9]{fig8}
  \caption{The $1/L$ dependences of the crossing temperature $T_L$ of the 
           excess free-energies for the lattice sizes of $L$ and $L+1$.}
  \label{fig:8}
\end{figure}


\begin{thebibliography}{999}
\bibitem{rf:22} H. Kawamura: private communication.
He insists that ${\theta}_{\rm rot}$ could not be larger than 
${\theta}_{\rm rev}$, because any rotation can be made from two 
sebsequent reversals.
\end{thebibliography}
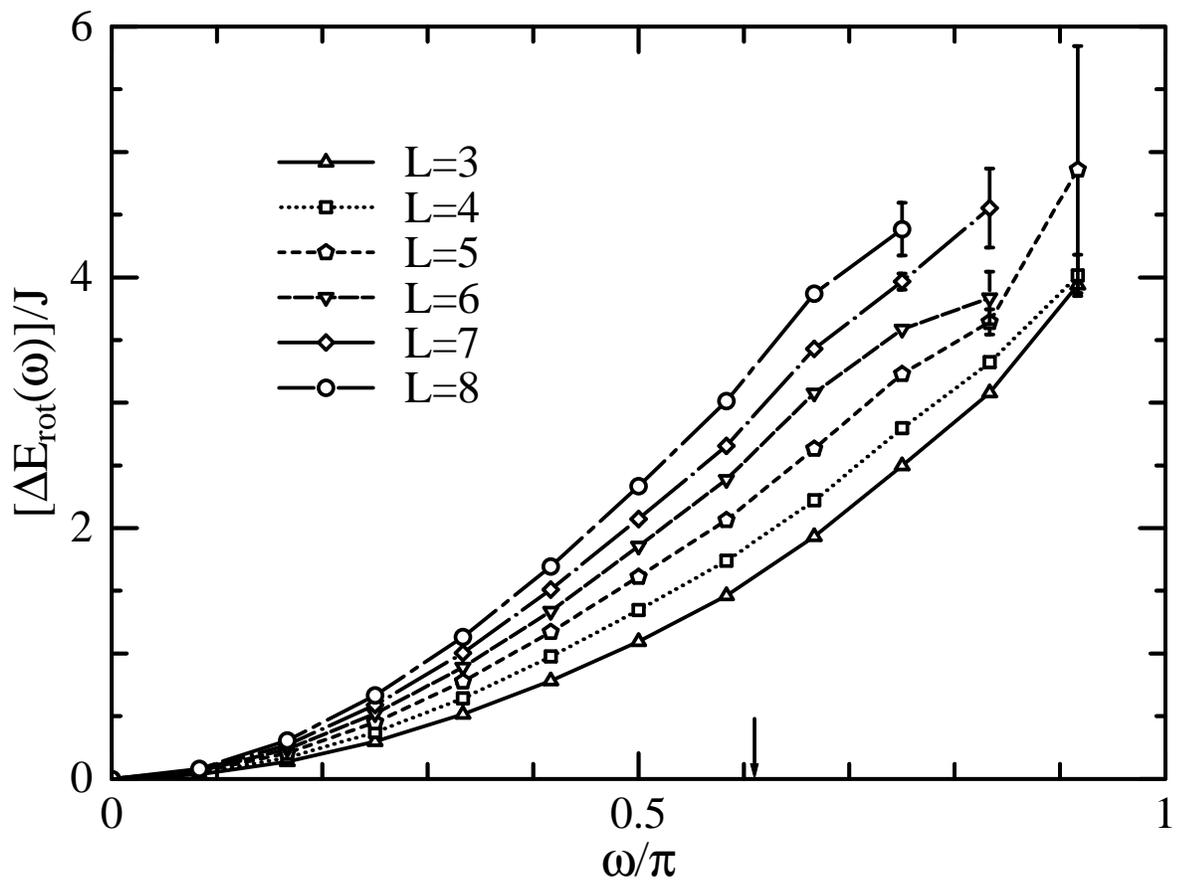
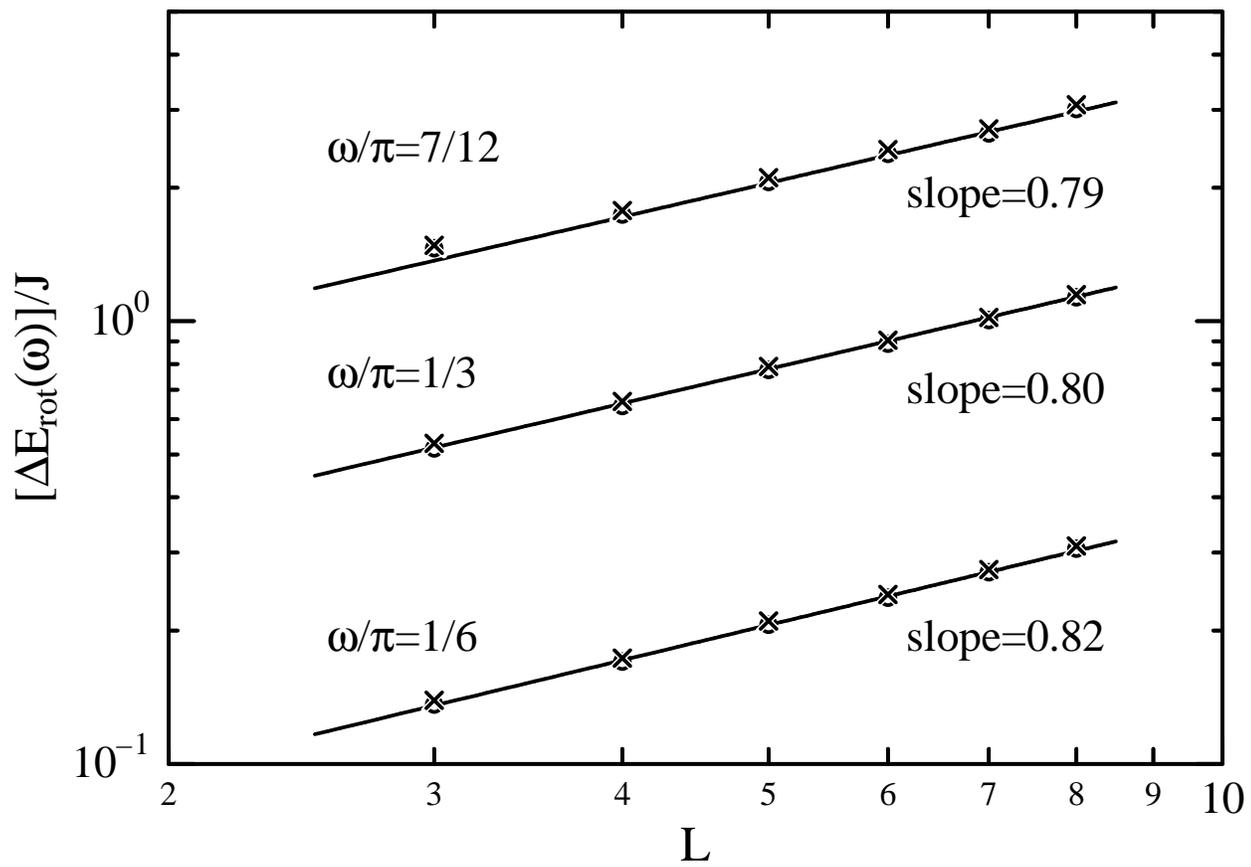
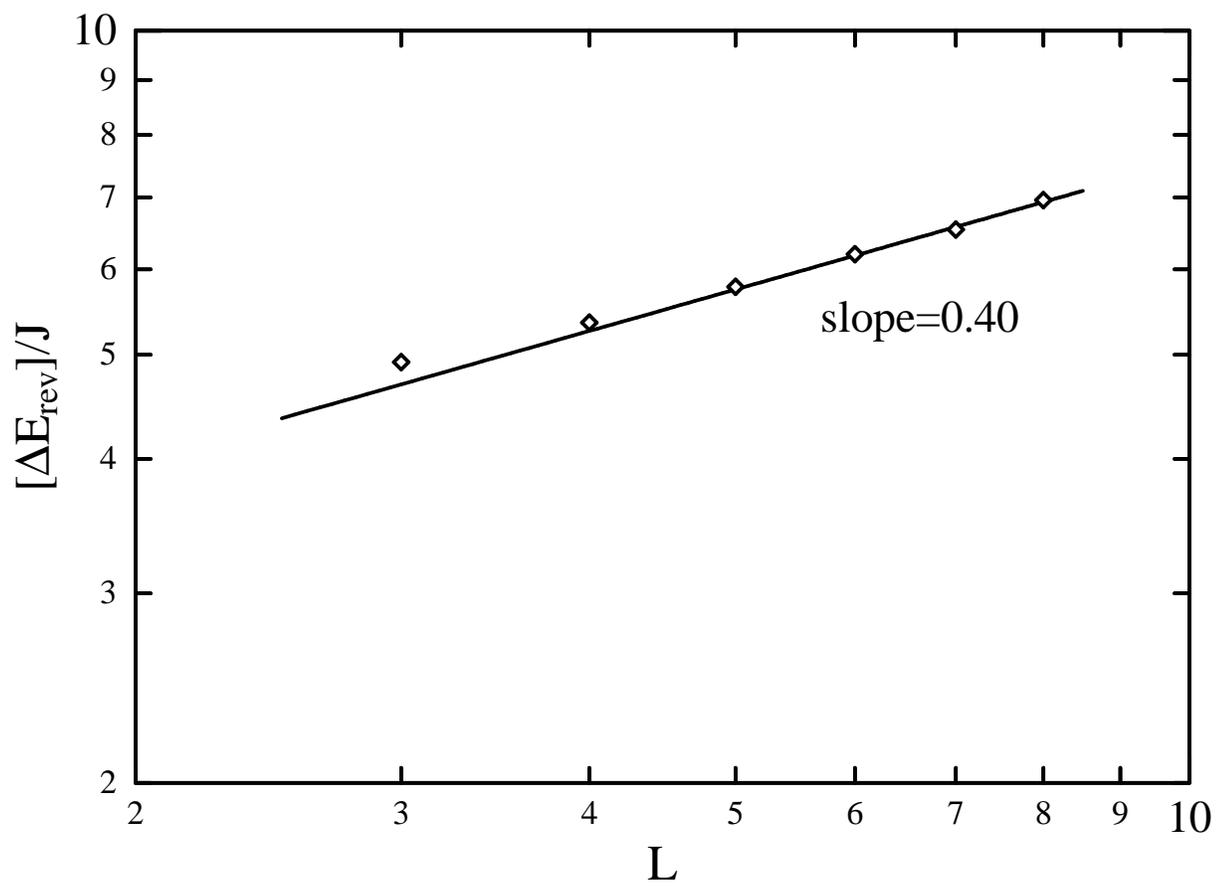
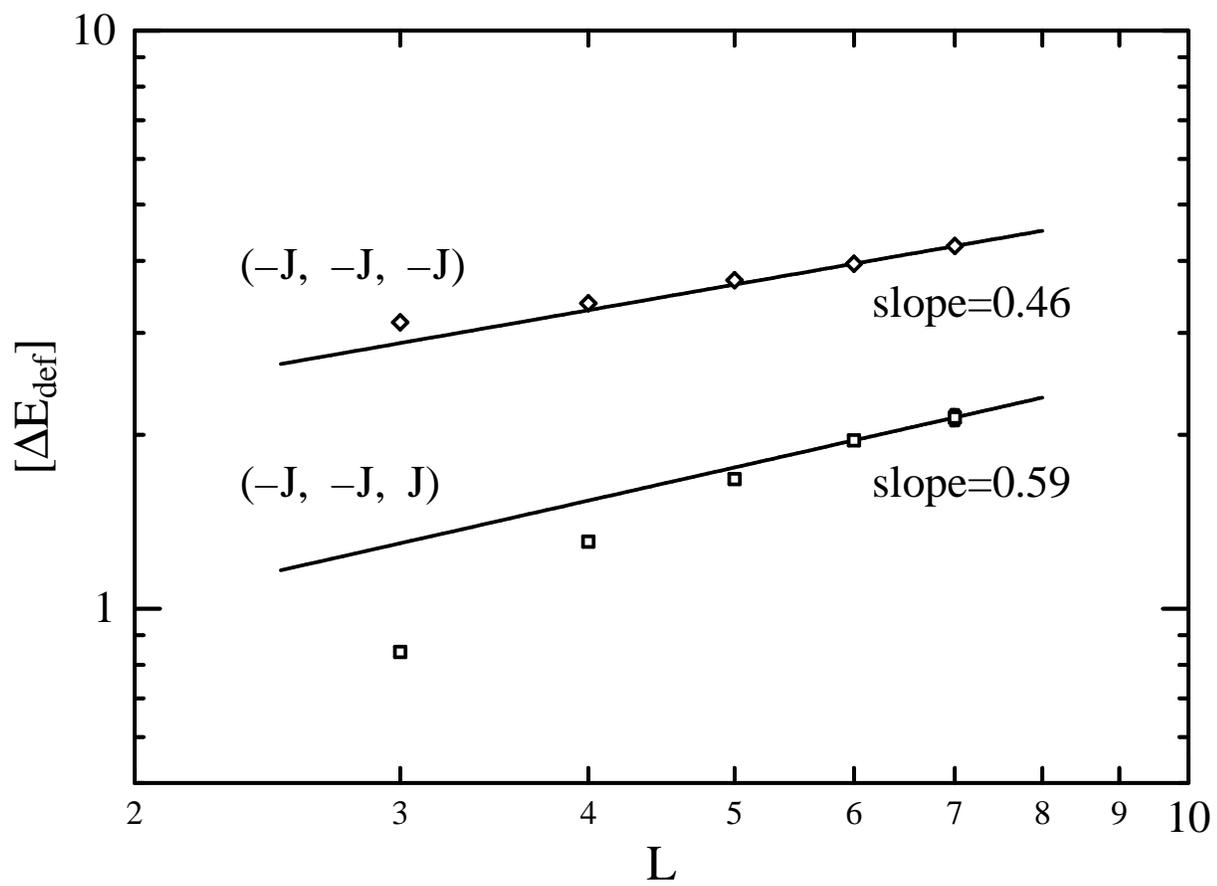
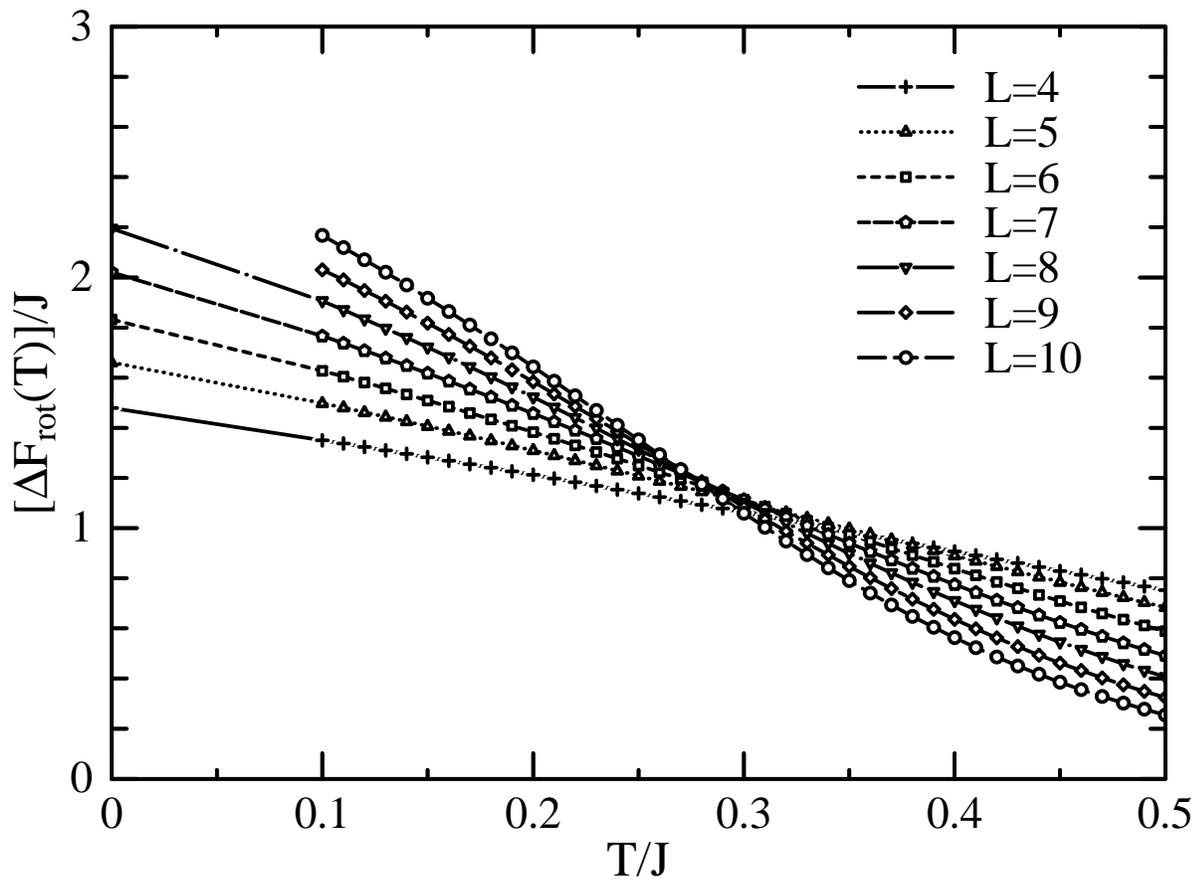
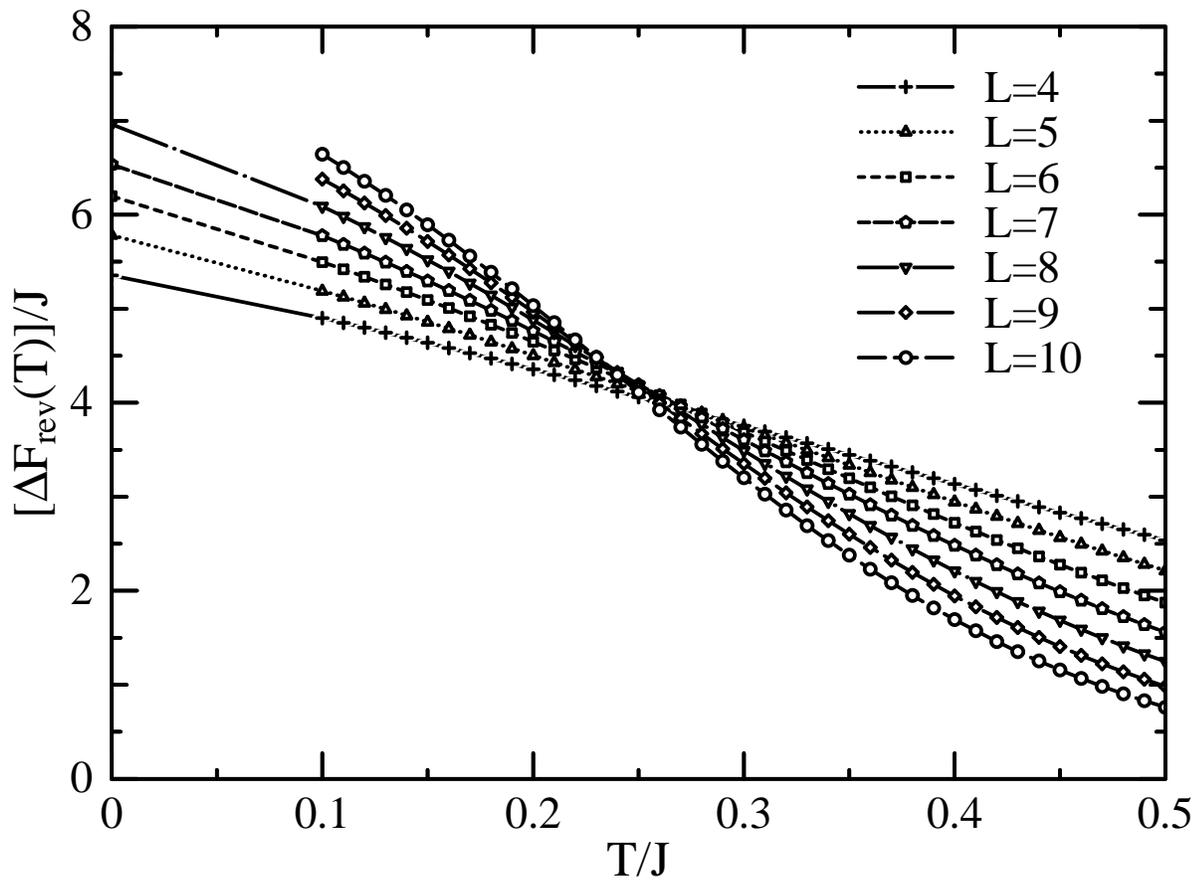
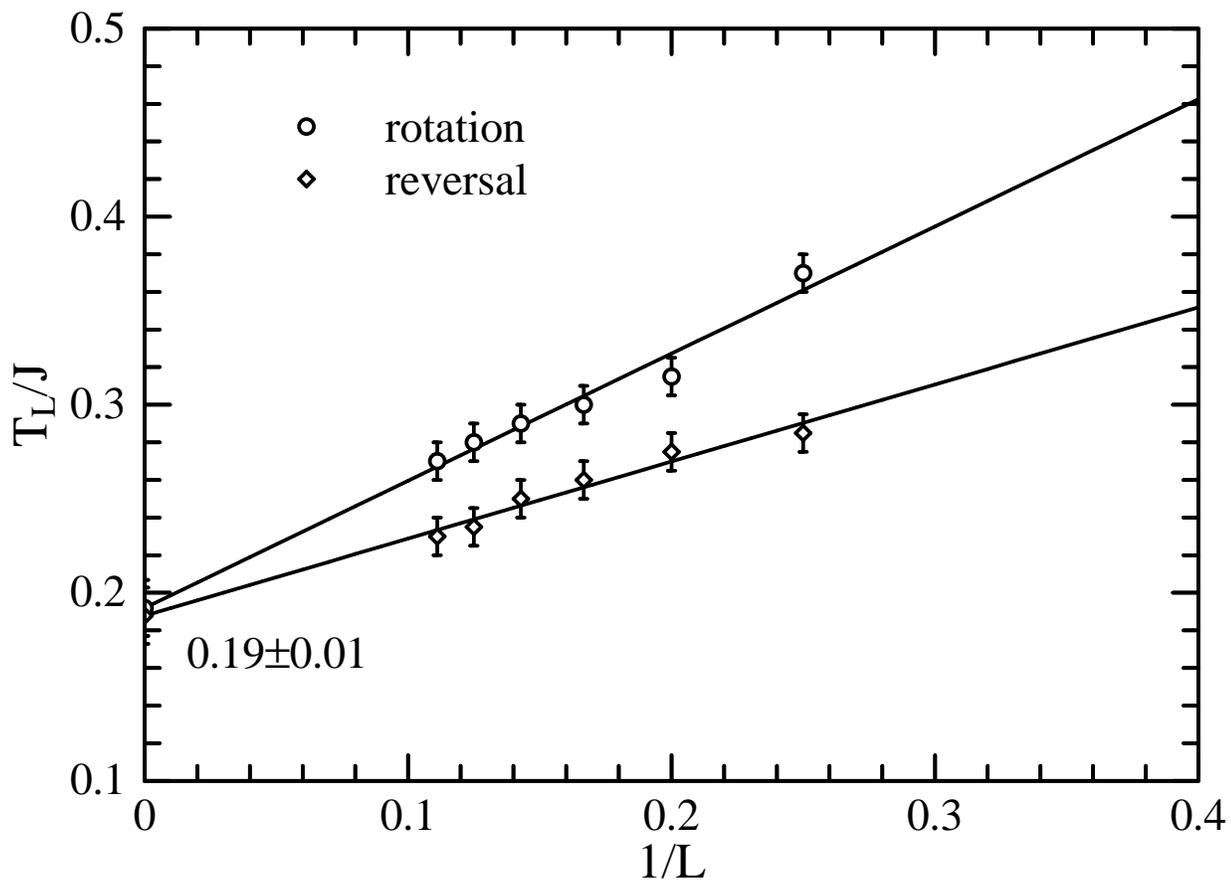
\end{document}